\documentclass[prl,aps,twocolumn,amsfonts,showpacs,superscriptaddress,floatfix]{revtex4}

\usepackage{color}
\usepackage{graphicx}
\usepackage[tight]{subfigure}
\usepackage{multirow}
\usepackage{amsmath}

\begin{document}


\title{Magnetic Response in the Underdoped Cuprates}
\author{A. J. A. James}
\affiliation{Condensed Matter Physics and Material Science Department, Brookhaven National Laboratory, Upton, NY 11973}
\author{R. M. Konik}
\affiliation{Condensed Matter Physics and Material Science Department, Brookhaven National Laboratory, Upton, NY 11973}
\author{T. M. Rice}
\affiliation{Condensed Matter Physics and Material Science Department, Brookhaven National Laboratory, Upton, NY 11973}
\affiliation{Institut f\"ur Theoretische Physik, ETH Z\"urich, CH-8093, Z\"urich, Switzerland}

\begin{abstract}
We examine the dynamical magnetic response of the underdoped cuprates by
employing a phenomenological theory of a doped resonant valence bond state where the Fermi
surface is truncated into four pockets.  This theory predicts a resonant spin response which with increasing
energy (0 to 100meV) appears as an hourglass.  The very low energy spin response is found at
$(\pi,\pi\pm\delta)$ and $(\pi\pm\delta,\pi )$ and is determined by scattering from the pockets' frontside 
to the tips of opposite pockets where a van Hove singularity resides.  At energies beyond 100 meV, strong scattering is seen
from $(\pi,0)$ to $(\pi ,\pi )$.
This theory thus provides a semi-quantitative description of the spin response seen in both INS and RIXS experiments 
at all relevant energy scales.
\end{abstract}

\maketitle

\newcommand{\del}{\partial}
\newcommand{\ep}{\epsilon}
\newcommand{\clsd}{c_{l\sig}^\dagger}
\newcommand{\cls}{c_{l\sig}}
\newcommand{\cesd}{c_{e\sig}^\dagger}
\newcommand{\ces}{c_{e\sig}}
\newcommand{\up}{\uparrow}
\newcommand{\down}{\downarrow}
\newcommand{\il}{\int^{\tilde{Q}}_Q d\la~}
\newcommand{\ilp}{\int^{\tilde{Q}}_Q d\la '}
\newcommand{\ik}{\int^{B}_{-D} dk~}
\newcommand{\ila}{\int d\la~}
\newcommand{\ilpa}{\int d\la '}
\newcommand{\ika}{\int dk~}
\newcommand{\tQ}{\tilde{Q}}
\newcommand{\rh}{\rho_{\rm bulk}}
\newcommand{\ri}{\rho^{\rm imp}}
\newcommand{\sh}{\sig_{\rm bulk}}
\newcommand{\si}{\sig^{\rm imp}}
\newcommand{\rph}{\rho_{p/h}}
\newcommand{\sph}{\sig_{p/h}}
\newcommand{\rp}{\rho_{p}}
\newcommand{\sip}{\sig_{p}}
\newcommand{\drph}{\delta\rho_{p/h}}
\newcommand{\dsph}{\delta\sig_{p/h}}
\newcommand{\drp}{\delta\rho_{p}}
\newcommand{\dsp}{\delta\sig_{p}}
\newcommand{\drh}{\delta\rho_{h}}
\newcommand{\dsh}{\delta\sig_{h}}
\newcommand{\enp}{\ep^+}
\newcommand{\enm}{\ep^-}
\newcommand{\enpm}{\ep^\pm}
\newcommand{\enph}{\ep^+_{\rm bulk}}
\newcommand{\enmh}{\ep^-_{\rm bulk}}
\newcommand{\enpi}{\ep^+_{\rm imp}}
\newcommand{\enmi}{\ep^-_{\rm imp}}
\newcommand{\enh}{\ep_{\rm bulk}}
\newcommand{\eni}{\ep_{\rm imp}}
\newcommand{\sig}{\sigma}
\newcommand{\la}{\lambda}
\newcommand{\ua}{\uparrow}
\newcommand{\da}{\downarrow}
\newcommand{\ed}{\epsilon_d}
{\bf Introduction:}
Neutron scattering studies of the magnetic properties of
underdoped cuprate superconductors have revealed an
unusual ‘hourglass’ pattern in the spin excitation spectrum
that persists into the normal state \cite{review1}. This spectrum
which is centered on $(\pi,\pi)$, can be divided into three
energy regions. At low energies the weight is shifted to
nearby incommensurate wavevectors, peaking along the
crystal axes. With increasing energy the weight is more
uniformly distributed around $(\pi,\pi)$ and contracts to form
a resonance centered on $(\pi,\pi)$. At still higher energies
a uniform ring appears evolving away from $(\pi,\pi)$.
Recent RIXS experiments \cite{rixs} have explored this high
energy region further.

A phenomenological theory for the underdoped
pseudogap phase by Yang, Rice and Zhang (YRZ) \cite{yrz}
has had considerable success in reproducing many
electronic properties \cite{yrz_review}.
In this letter we examine the spin
spectrum within this theory and show that key features of
the experiments at all three energies are reproduced. We
begin with a derivation of the YRZ ansatz starting from a t-
J model rather than from the overdoped Fermi liquid in the
original paper, following a recent suggestion by P. A. Lee \cite{pl_private}
which in turn is based on a small modification of Ref. \cite{ng}.
An RPA
form is used for the spin response similar to
that employed by Brinckman and Lee in their study of the
spin resonance in the superconducting state of overdoped
cuprates. The itinerant description in our approach differs from the often
used scenario of strong and slow stripe fluctuations which would show up as incommensurate
quasi-elastic peaks in the magnetic response \cite{stripes}.  We compare our results with data sets gathered
using both x-rays and neutrons.

\begin{figure}
\subfigure{\includegraphics[width=0.32\textwidth]{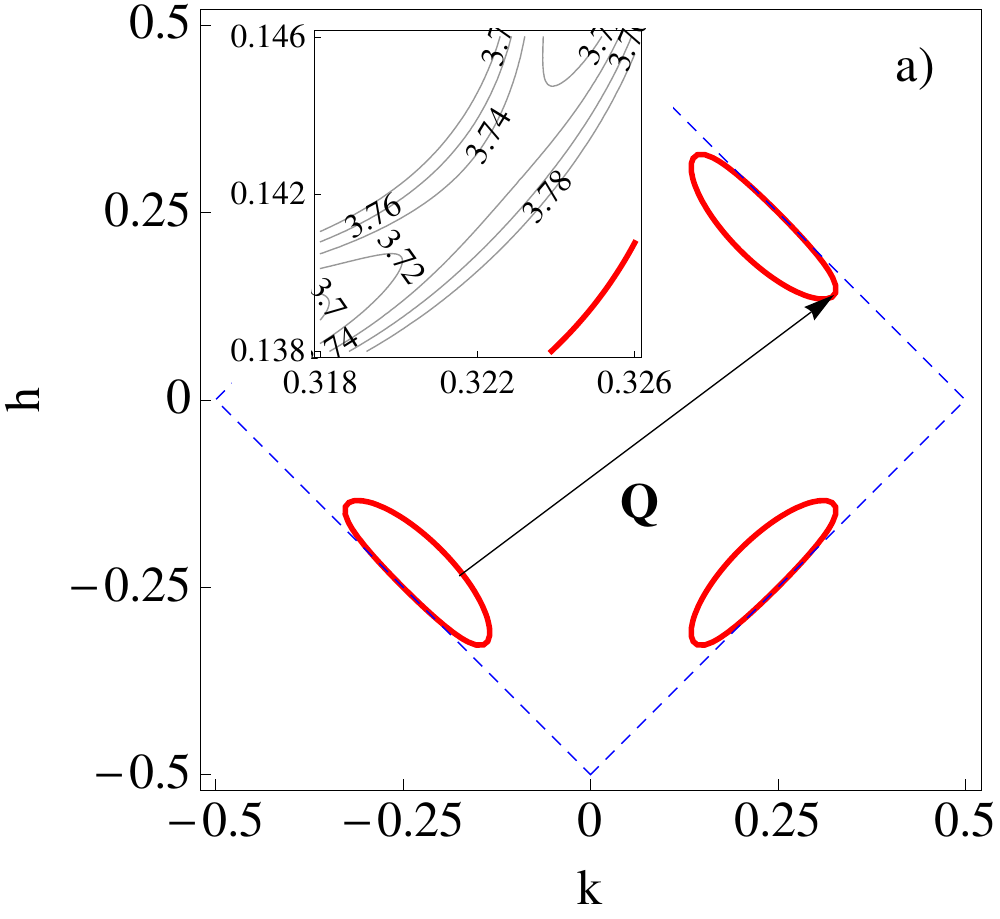}}%
\subfigure{\includegraphics[width=0.16\textwidth]{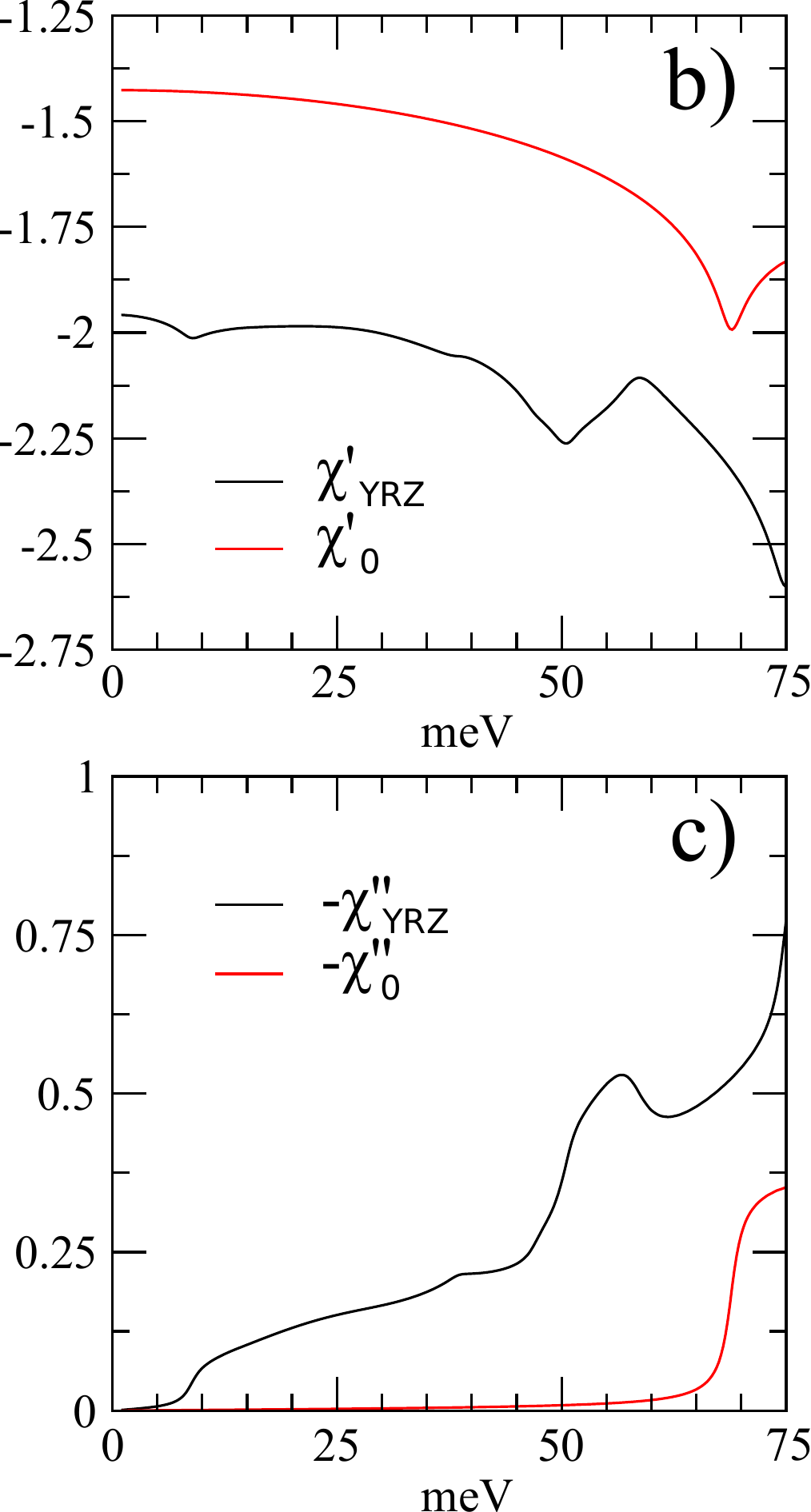}}%
\caption{a) The Fermi surface for hole doping, $x=0.12$. Hole pockets are marked in red (solid) while the 
lines of Luttinger zeros are blue (dashed). Also marked is a nesting vector ${\bf Q}=(0.5,0.375)$ (in reciprocal lattice units), 
connecting the tip of a pocket to the frontside of another pocket. Inset: A plot demonstrating that at the
tip of the pocket there
is a saddlepoint in the superconducting quasi-particle dispersion and
hence a van Hove singularity. Contours are labelled in meV. The parameters used
here are $t(x)=70$meV, $t'(x)=-0.18t(x)$, $t''(x)=0.12t(x)$, $\Delta_0=0.34t(x)$
and $\Delta_{SC}=0.05t(x)$.
b) The real parts of $\chi_{YRZ}({\bf Q},\omega)$ and $\chi_0({\bf Q},\omega)$ vs. $\omega$. c) The imaginary parts.}
\label{fig:fs}
\end{figure}

{\bf YRZ Spin Response:}
The YRZ ansatz, as originally conceived, was an ansatz for the single particle Green's function (GF)
of the underdoped cuprates.  The Fermi surface associated with it is truncated and composed of four nodal
pockets (see Fig. 1).  The area of the pockets is proportional to the level of doping, x.
This GF is also characterized by lines of Luttinger zeros which coincide with the magnetic Brillouin zone (BZ)
or Umklapp surface \cite{honer}
(see Fig. 1).   
The ansatz was inspired by an analysis of a system of weakly coupled Hubbard ladders where a similar
phenomenology was found to hold \cite{krt}.

To extend the YRZ ansatz from the single particle GF to the spin response, we first elucidate the connection between YRZ and
the slave boson (SB) treatment of the t-J Hamiltonian.  SBs provide a natural RPA-like form to the spin response and we 
intend to adapt this to the assumptions of YRZ.
For this purpose we then write the $t-J$ Hamiltonian as
\begin{eqnarray}
H &=& -\sum_{ij\sigma}t^{nn}_{ij} c^\dagger_{i\sigma}c_{j\sigma} -\sum_{ij\sigma}t^{nnn}_{ij} c^\dagger_{i\sigma}c_{j\sigma} 
+ \frac{1}{2} \sum_{ij}J_HS_i\cdot S_j\cr\cr
&\equiv & H_t^{nn} + H_t^{nnn} + H_{J_H}
\end{eqnarray}
The Hamiltonian is divided into terms involving nearest neighbour (NN) hopping, $H_t^{nn}$, terms
involving next nearest neighbour (NNN) hopping (and beyond), $H_t^{nnn}$, and a spin-spin interaction, $H_{J_H}$.  
We choose this separation because of
the focus the YRZ ansatz places upon the Luttinger zeros found at the magnetic BZ.  
The nearest neighbour dispersion, $\xi_{0}(k)=-2t(x)(\cos k_x + \cos k_y)$ (for the definition of t(x) see \cite{yrz,supp_mat})
vanishes on this line while
that of the NNN hopping does not. (We show in the supplementary material 
\cite{supp_mat} that one can derive a YRZ-like propagator
while treating NN and NNN hopping on the same footing.)
We now subject $H_t^{nn} + H_{J_H}$ to the standard slave boson mean field treatment (leaving $H_t^{nnn}$ to later).  We thus factor
the fermions, $c^\dagger_{i\sigma}$, into spinons, $f^\dagger_{i\sigma}$ and holons, $b_i$ via
$
c^\dagger_{i\sigma} = f^\dagger_{i\sigma}b_i,
$
where the spinons and holons are subject to the constraint
$
\sum_\sigma f^\dagger_{i\sigma}f_{i\sigma} + b^\dagger_ib_i = 1.
$
At this level the spinon Green's function can be shown to be \cite{bl}
\begin{align}
G^f_{\sigma}(\omega,{\bf k}) = 
\frac{1}{\omega - \xi_0({\bf k}) - \Sigma_R(\omega,{\bf k})}, 
\end{align}
where $\Sigma_R = |\Delta_R({\bf k})|^2/(\omega + \xi_0({\bf k}))$ and 
$\Delta_R({\bf k}) = \Delta_0(x)(\cos k_x - \cos k_y)$.  Here $t(x)$ and $\Delta_0(x)$ are doping dependent parameters.
The single particle Green's function, $G^c_{\sigma}$, is given
directly in terms of the spinon Green's function because we assume the bosons are nearly condensed and so
replace the boson propagator $\langle b_i^\dagger(\tau )b_j(0)\rangle$ by $g_t(x)$:
$G^c_{\sigma}(\omega,{\bf k})= g_t(x) G^f_{\sigma}(\omega,{\bf k})  $
(in slave boson mean field $g_t(x)=x$ \cite{bl}; in the Gutzwiller approximation $g_t(x)=2x/(1+x)$ \cite{yrz}).
This is close to the YRZ form but differs in that the full dispersion in the denominator is replaced by the dispersion due to NN hopping. 

To bridge the gap between the SBMFT and YRZ, we then turn to the so far neglected NNN hopping, $H_t^{nnn}$.  
Treating this term in mean field theory (MFT) moves the Luttinger zeros off the magnetic Brillouin zone and so
we instead use an RPA like approximation (see Fig. 2) leading to
\begin{align}\label{renorm_spin_prop}
G^f_{\sigma}(\omega,{\bf k})  &=
\frac{1}{\omega - \xi_0({\bf k}) - \xi'({\bf k}) - \Sigma_R(\omega,{\bf k})}.
\end{align}
Here $\xi'(k)= -4t'(x)\cos k_x\cos k_y -2t''(x)(\cos 2k_x + \cos 2k_y)$ is the
dispersion due to the NNN terms.  The spinon propagator in this form now immediately gives the YRZ ansatz.
\begin{figure}
\includegraphics[width=0.34\textwidth]{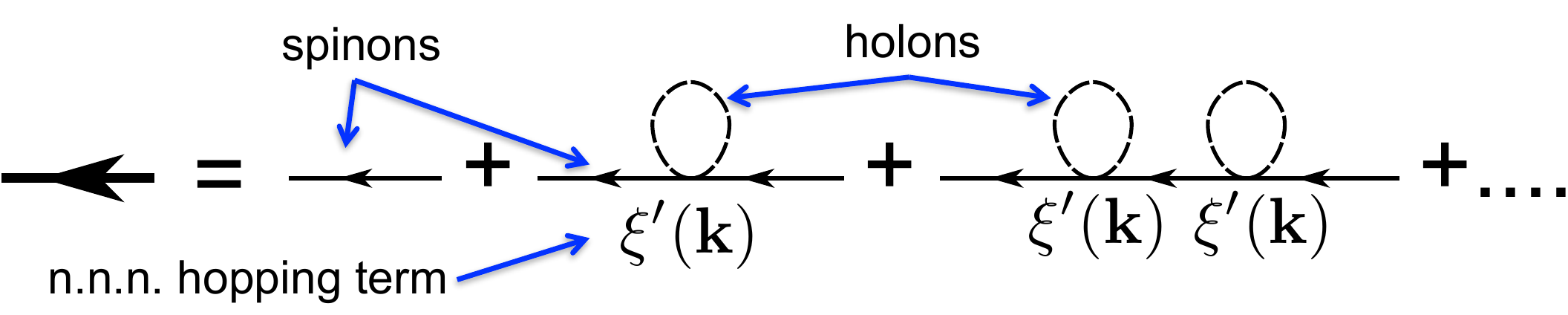}
\caption{RPA form of the YRZ spinon propagator in terms of SB propagators.}
\label{fig:fs}
\end{figure}
An important consequence of the non-MFT treatment of the $H^{nnn}_t$ is that 
spinons and holons are bound together.
This binding distinguishes the YRZ ansatz from the standard mean field SB approximation which leads to an 
expanded Hilbert space with independent spinons and holons.  
A second consequence is the absence of an anomalous spinon propagator (or at least the coherent
part thereof), consistent with an underlying RVB assumption that
spin correlations are short- not long-ranged in the YRZ ansatz.

This form (Eqn. \ref{renorm_spin_prop})
applies in the normal phase and can be generalized to the d-wave superconducting state, e.g. see \cite{yrz_review}.
Note that YRZ gives a two-gap description of the pseudogap phase with a separate RVB ($\Delta_0$) and pairing
$\Delta_{SC}$ gaps.

We now turn to the YRZ spin response.  In slave bosons, neglecting the effects
of NNN hopping, the spin response naturally takes on an 
RPA-like form \cite{bl}:
\begin{equation}
S(\omega,{\bf k}) = -\frac{3}{\pi}\mathrm{Im}\frac{\chi_0(\omega,{\bf k})}{1-J({\bf k})\chi_0(\omega,{\bf k})}.
\end{equation}
Here $\chi_0(\omega,{\bf k})$ is the bare particle-hole bubble for the spinons (including anomalous
contributions)
and $J({\bf{k}})= J(\cos k_x + \cos k_y)$.

How now does our non-mean field treatment of $H^{nnn}_t$ alter this?  Its effect is two-fold.
Firstly we no
longer include a contribution to $\chi_0$ from the anomalous spinon Green's functions.  
And to determine how $t^{nnn}$ dresses the normal spinon Green's functions,
we employ the same approximation that led to the YRZ ansatz.
Namely we only allow diagrams involving vertices where the boson lines of the vertex are tied together.  Under such a restriction,
$t^{nnn}$ only dresses the individual spinon propagators making up the particle-hole bubble entering
$\chi_0$.
In this fashion the YRZ spin ansatz takes the form
\begin{equation}
S_{YRZ} (\omega,{\bf k}) = -\frac{3}{\pi}\mathrm{Im}\frac{\chi_{YRZ}(\omega,{\bf k})}{1-J({\bf k})\chi_{YRZ}(\omega,{\bf k})},
\end{equation}
where $\chi_{YRZ}$ is simply a particle-hole bubble made up of YRZ quasi-particles.

In computing the spin response we treat $J$ as a fitting parameter for each doping (and different from $J_H$).  
We do not expect 
the underlying mean field treatment
to accurately treat the renormalization of $J$ which is inevitably doping dependent.  In particular in the presence of strong
scattering connecting the magnetic Brillouin zone boundaries, we expect
$J$ to be strongly modified.  This is not merely a feature of the YRZ theory but is generic to slave boson flavoured theories.
In Ref. \cite{bl}, $J$ had to be sharply reduced in order to produce an AF ordering transition at approximately
the correct doping.

{\bf Results:}
We begin with the lower energy ($\omega<100$meV) spin response in the underdoped cuprates which has a 
universal hour glass shape \cite{review1,lsco8.5,lsco16a,lsco16b}
with strong incommensurate response at low energies (i.e. $\omega \approx 2\Delta_{SC}$)
concentrated at four points,
$(\pi,\pi\pm\delta)$ and $(\pi\pm\delta,\pi)$.  As energy is initially increased this response evolves inwards
towards $(\pi,\pi)$ and simultaneously becomes more isotropic in its distribution about ($\pi, \pi$).  Whether
this inward dispersion reaches $(\pi ,\pi)$ is a function of the particular cuprate being examined.  
As the
energy is further increased this inward evolution is reversed and the response moves outwards from ($\pi , \pi$).
With this outward dispersion, the response is more isotropically 
distributed about $(\pi ,\pi )$.

\begin{figure}[th]%
\centering%
\subfigure{\includegraphics[width=0.2\textwidth]{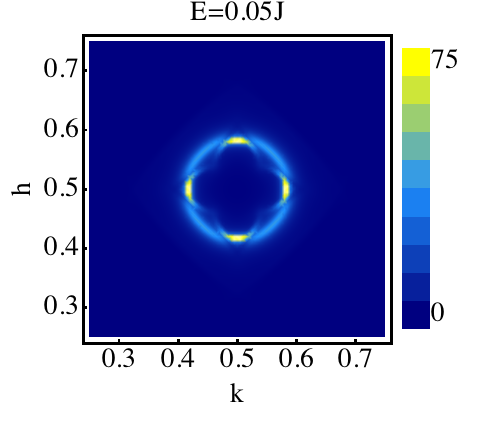}}%
\subfigure{\includegraphics[width=0.2\textwidth]{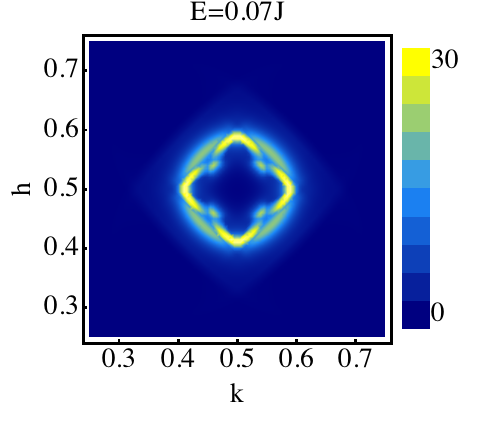}}\\%
\vskip -.15in
\subfigure{\includegraphics[width=0.2\textwidth]{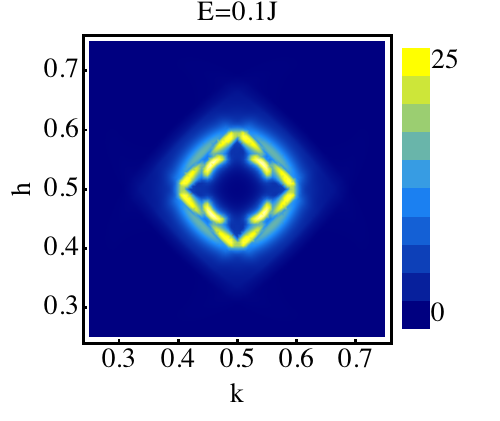}}%
\subfigure{\includegraphics[width=0.2\textwidth]{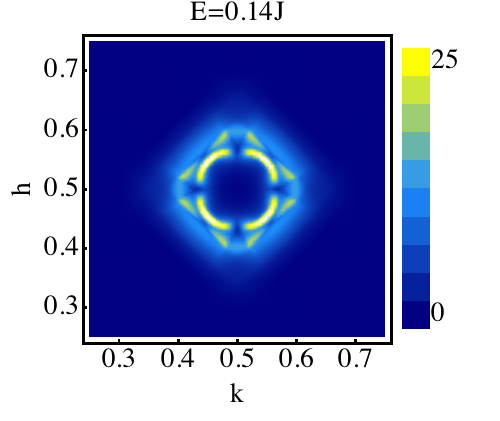}}\\%
\vskip -.15in
\subfigure{\includegraphics[width=0.2\textwidth]{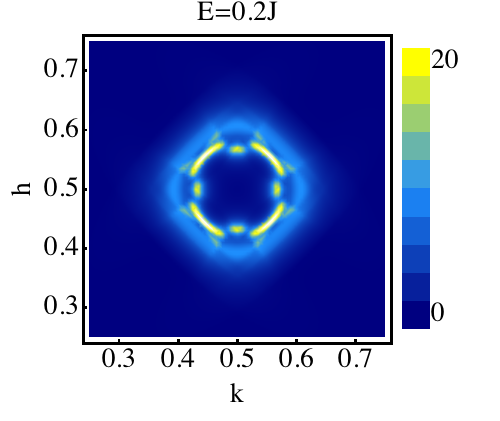}}%
\subfigure{\includegraphics[width=0.2\textwidth]{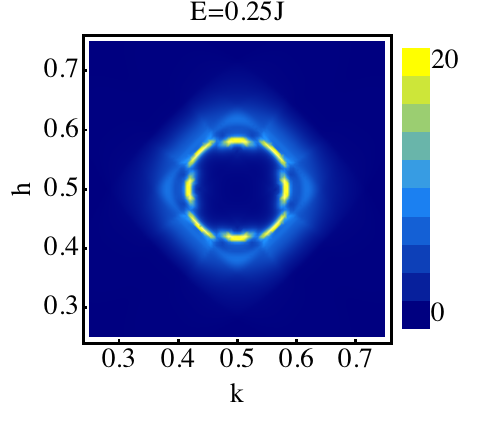}}\\%
\vskip -.15in
\subfigure{\includegraphics[width=0.2\textwidth]{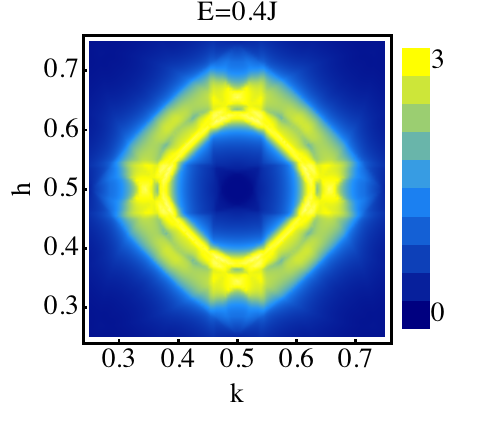}}%
\subfigure{\includegraphics[width=0.2\textwidth]{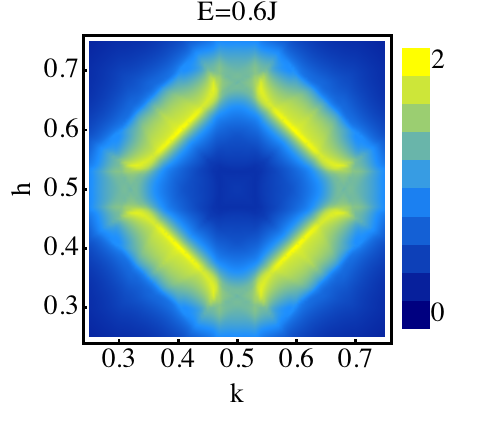}}%
\vskip -.15in
\caption{Constant energy slices of the spin response
for $x=0.12$ in the SC phase -- the parameters used are the same as
listed in Fig. 1, with $J=0.14$meV for our theory.}%
\end{figure}%

We see these general features in the constant energy scans of the q-dependent spin response presented in Fig. 3
for the superconducting case.
In this figure we have chosen parameters appropriate for the description of underdoped $\rm La_{2-x}Sr_xCuO_4$.  
We see at very low energies
($0.05J$) the primary response is at $(\pi,\pi\pm\delta)$ and $(\pi\pm\delta,\pi)$ with $\delta=0.16\pi$.  
As the energy is increased
there is a slight inward dispersion ($\delta$ decreases slightly) and the spin response is found circularly distributed
about $(\pi ,\pi)$.  This dispersion reverses at $\omega\sim 0.2J$ and begins to move outwards.  In this
energy range the greatest response is found about $(\pi\pm\delta' ,\pi\pm\delta')$.

The response found at $(\pi,\pi\pm\delta)$ and $(\pi\pm\delta,\pi)$ at $0.05J$meV can be directly ascribed
to transitions between the fronts of the pockets and the tips of opposite pockets (see the vector $Q$ in Fig. 1).
In general the presence of the pockets in the YRZ theory allows for low 
energy scattering in a larger portion of the Brillouin
zone than in theories where the spinon Fermi surface consists of four points 
coinciding with nodes of the SC order parameter
(see Figs. 1b and 1c for a comparison of $\chi_{YRZ}$ and $\chi_{0}$; $\chi_0$ 
is the bare particle-hole bubble for the standard
slave boson description of the spin response \cite{bl}).
Moreover in the presence of a SC gap, the tips of the pockets see a saddle point in dispersion with a corresponding
van Hove singularity thus further enhancing the low energy scattering.

\begin{figure}[th]
\centering
\subfigure{\includegraphics[width=0.18\textwidth]{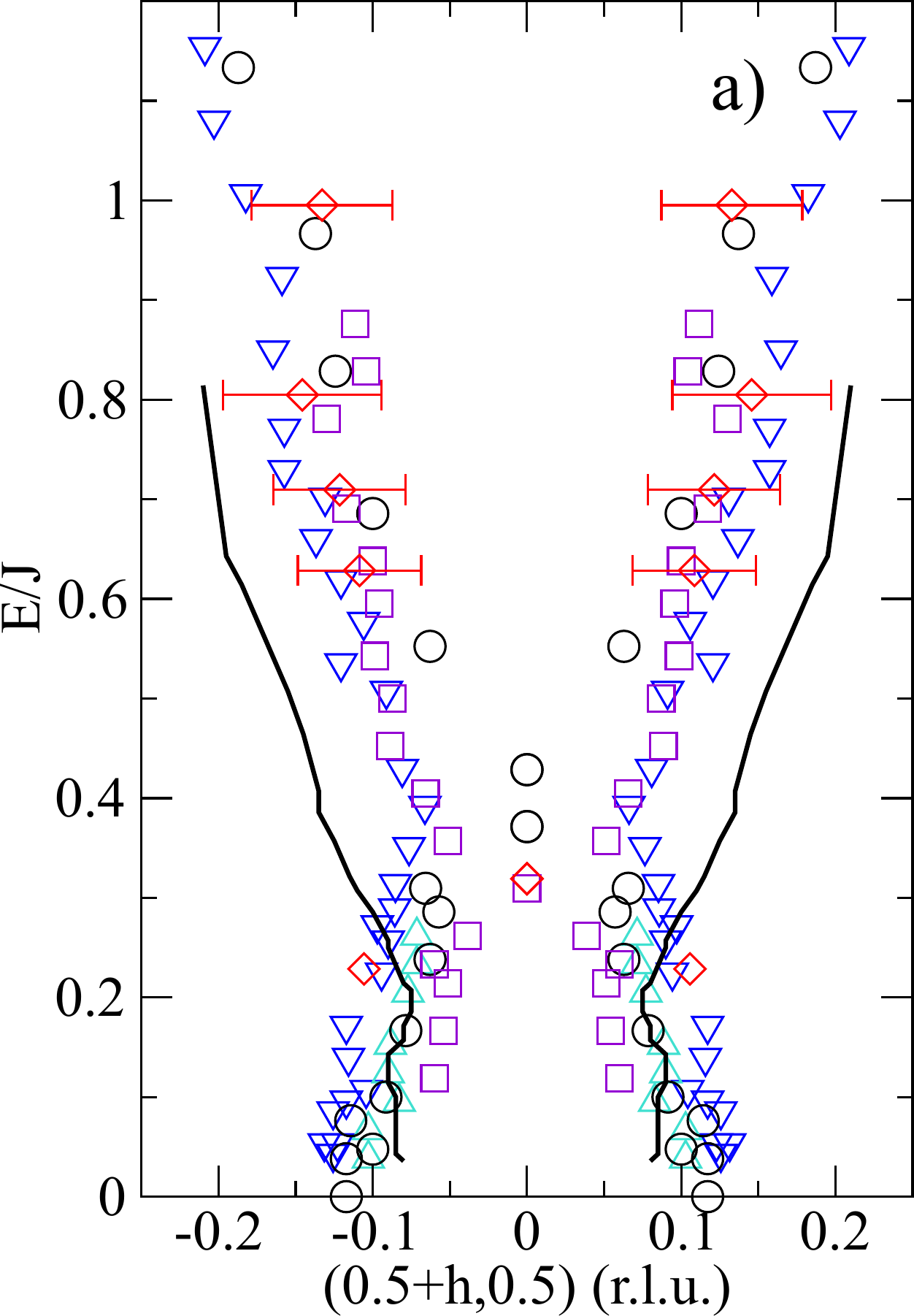}}
\subfigure{\includegraphics[width=0.25\textwidth]{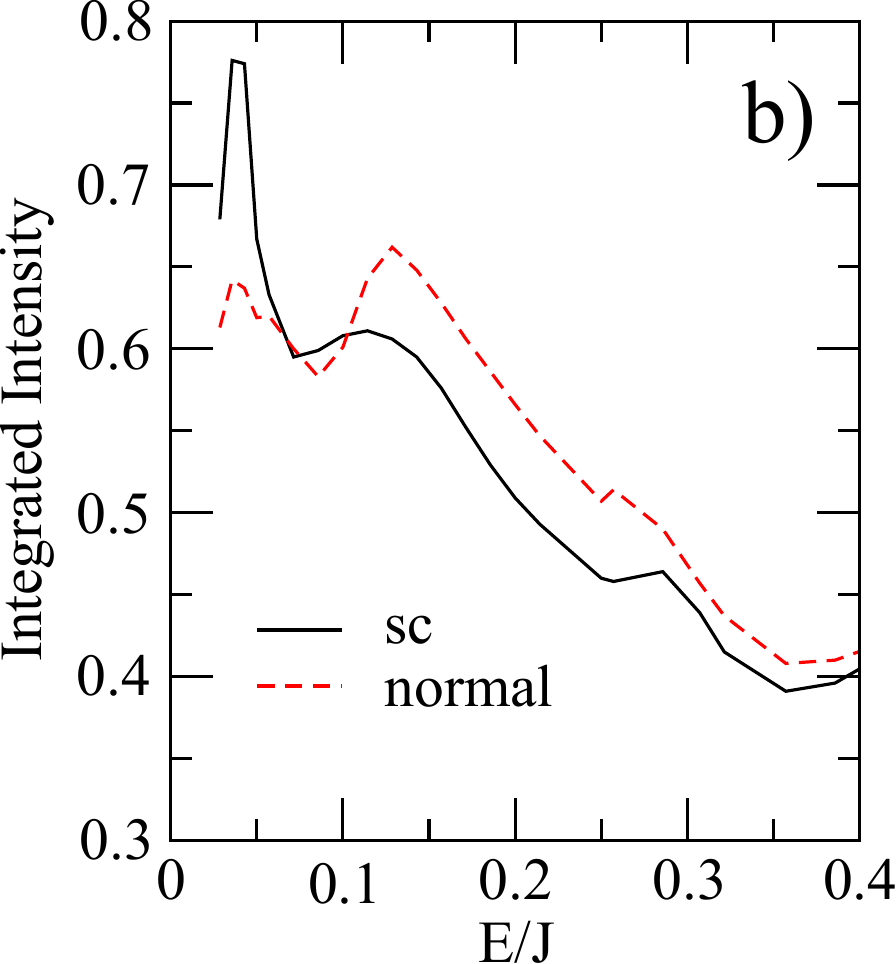}}
\caption{a) Hourglass dispersion of the resonance near $\pi,\pi$. The thick black line is the 
position of the maximum 
intensity peak after integrating the numerical data over a strip of width $2\pi/25$ 
along the parallel direction, averaged over 
sections of length $2\pi/33$. Experimental data points (appropriately rescaled)
are taken from \cite{review,review1}:
`up' triangles, 
La$_{1.90}$Sr$_{0.10}$CuO$_4$ 
\cite{lsco8.5}; 
circles, La$_{1.875}$Ba$_{0.125}$CuO$_4$ \cite{jtrannature}; `down' triangles,  La$_{1.84}$Sr$_{0.16}$CuO$_4$ 
\cite{lsco16b}; 
squares,  YBa$_{2}$Cu$_{3}$O$_{6.5}$ \cite{stock} and diamonds, YBa$_{2}$Cu$_{3}$O$_{6.6}$ {hayden}. 
b) k-integrated spin response with and without a superconducting gap.}
\end{figure}

With increasing energy, the $\bf k$-points with maximal intensity move inward towards $(\pi ,\pi )$, 
albeit in an uneven fashion (there is a sudden movement inward at $0.125J$) while at the same time becoming
more isotropically distributed about $(\pi ,\pi )$.  This behavior is shared not only by underdoped
$\rm La_{2-x}Sr_xCuO_4$ \cite{lsco8.5} but also its optimally doped counterpart \cite{lsco16a,lsco16b}.
It is also seen in stripe stabilized $\rm La_{2-x}Ba_xCuO_4$ \cite{jtrannature} and $\rm YBCO$ 
\cite{review,review1}.  At an energy between $0.4J$ and $0.5J$, the point of maximal intensity begins to drift
outward from $(\pi,\pi )$, again a universal feature of the magnetic response in the cuprates.
We explicitly plot in Fig. 4a the evolution of the k-point of maximal intensity as a function of energy,
comparing its evolution with a number of cuprates.

In the normal state, low energy spectral weight is found not just in directions parallel to the
crystal axes but in the nodal directions as well (see Fig. 5).  
While parallel scattering still dominates at low energies, the response
is less concentrated in such areas and weight does appear
along the nodal directions (at least in the LSCO family) \cite{lsco8.5,lsco16a}.  
\begin{figure}[th]%
\centering%
\subfigure{\includegraphics[width=0.2\textwidth]{x12_7meV.png}}%
\subfigure{\includegraphics[width=0.2\textwidth]{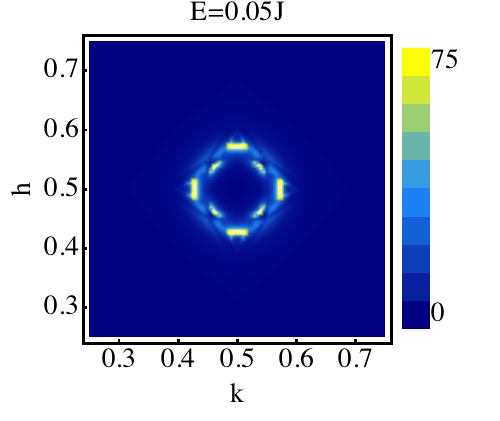}}%
\vskip -.15in
\caption{$x=0.12$ and $\omega=7$meV constant energy slices for the SC phase (left)
and the normal phase (right).}
\end{figure}
 
Underlying our calculations of the magnetic response
is the assumption that itinerant quasi-particles (even if
heavily dressed) can explain this response in the cuprates.
While there is evidence that at least part
of the spin response must be ascribed to localized spins
\cite{jtrannature,bscco}, there is also evidence that impurities introduce local spins,
e.g. Zn doped into YBCO \cite{ZnYBCO}
and earlier studies. The full
cuprate magnetic response requires a mixture of the two. 
However one experimental feature of the spin response that points to itinerant quasi-particles is 
the depression of the
k-integrated spin response at $\omega < 2 \Delta_{SC}$ upon decreasing $T<T_c$. 
This behavior is seen in both
the LSCO \cite{lsco8.5,lsco16a,lsco16b} and $\rm YBCO$ families and we see it as well
in our calculations (Fig. 4b).
We also see in Fig. 4b that our calculated integrated
intensity has a two peak structure, with one peak at energies
close to $0.05J$ and one at energies at $\approx   0.12J$.  This
doubling of peaks is seen in near optimally doped LSCO
\cite{lsco16a,lsco16b}. In underdoped LSCO at least the lower energy
peak has been observed \cite{lsco8.5}.

Turning to high energies, $\omega > 100$meV, we find the YRZ spin response is also able to 
explain key features in the spin response recently measured by RIXS experiments.
In Fig. 6 we plot the spin response for energies $100{\rm meV} < \omega < 300{\rm meV}$ for two cuts 
in the Brillouin zone. We see two features emanating from $(0,0)$. One disperses towards $(\pi,0)$ as energy
is increased (corresponding well with the reported
paramagnon-like excitation in the RIXS data of
\cite{rixs} on a variety of cuprates). The other, with a
considerably greater spin velocity, evolves towards $(\pi,\pi)$.
This dispersing paramagnon excitation naturally
appears from a two-band factorization of YRZ where the propagator can be written in the form \cite{yrz_review}
\begin{equation}
G^c_{\sigma} (\omega, {\bf k}) 
= \frac{z_+({\bf k})}{\omega-\omega_+({\bf k})} + \frac{z_-({\bf k})}{\omega-\omega_-({\bf k})}.
\end{equation}
The paramagnon results from a particle-hole excitation from the lower band, $\omega_-({\bf k})$,
to the upper band, $\omega_+({\bf k})$.

\begin{figure}[th]
\centering
\subfigure{\includegraphics[width=0.4\textwidth]{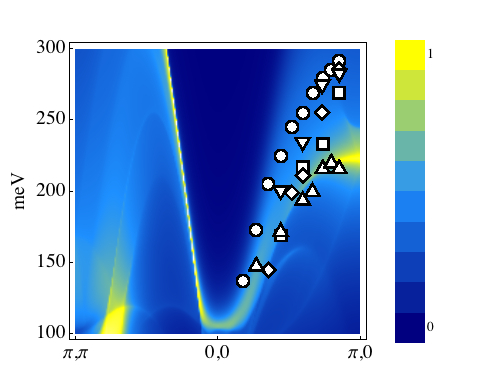}}
\caption{The spin response for energies from 100meV to 300meV, for cuts from $(\pi,\pi)$ to $(0,0)$ to $(\pi,0)$ 
in the Brillouin zone (same choice of parameters as previously). Also plotted are data points from \cite{rixs}: circles, Nd$_{1.2}$Ba$_{1.8}$Cu$_{3}$O$_{6}$; squares, YBa$_{2}$Cu$_{3}$O$_{7}$; diamonds, Nd$_{1.2}$Ba$_{1.8}$Cu$_{3}$O$_{7}$; `up' triangles, YBa$_{2}$Cu$_{4}$O$_{8}$; `down' triangles, YBa$_{2}$Cu$_{3}$O$_{6.6}$.}
\end{figure}

In conclusion we have shown that calculations of the magnetic response based upon itinerant 
YRZ quasi-particles reproduces
key features reported in experiments on the spin response of the underdoped cuprates at both 
low and high energies in a satisfactory way.

The authors acknowledge support from the CES, an ERFC funded by the DOE's OBES.
We also thank J. Tranquada, J. Hill, and M. Dean for useful conversations.

\break 

\appendix

\section{Supplementary material}

In this appendix we give an alternate derivation of YRZ from SBMFT that does not depend on 
treating the nearest neighbour hopping term in the Hamiltonian, $H^{nn}_t$, on a different
footing than $H^{nnn}_t$.

Here we instead divide the full Hamiltonian into its Heisenberg piece, $H_{J_H}$, and its hopping
terms, $H_t^{nn}+H^{nnn}_t$.  We first focus on treating $H_{J_H}$ in SBMFT.  Using standard SBMFT,
we find that the spinon propagator is now given by
\begin{equation}
G^f_\sigma (\omega,{\bf k}) =  
\frac{1}{\omega - \tilde\xi_0({\bf k}) 
- \frac{|\Delta_R({\bf k})|^2}{\omega+\tilde\xi_0({\bf k})}},
\end{equation}
where 
$$
\tilde\xi_0({\bf k}) = -2\tilde t(x)(\cos k_x + \cos k_y).
$$
$\tilde\xi_0$ differs from $\xi_0$ in that $\tilde t(x) = 3/8g_s(x)J_H\chi$ instead of
$t(x) = g_t(x)t_0+3/8g_s(x)J_H\chi$ where $t_0$ is the bare nearest neighbour hopping strength,
$J_H$ is the bare Heisenberg coupling, $g_s(x)$ is the amount $J_H$ is renormalized in the Gutzwiller
projection, and $\chi=0.338$.  We note that at
zero doping ($x=0$), this propagator coincides with the SB propagator in Eqn. (2).

To treat the remaining hopping terms, $H_t^{nn}+H^{nnn}_t$, we now proceed as we did with $H^{nnn}_t$ alone
in the main text. Using an RPA approximation, we find for this version of the YRZ propagator
\begin{eqnarray}\label{renorm_spin_prop}
G^f_\sigma (\omega,{\bf k}) &=& 
\sum_{n=0}^\infty (\tilde\xi'({\bf k}))^{n}
G^f_\sigma |_{\xi'({\bf k})=0} (\omega,{\bf k}))^{n+1}\cr
&=& 
\frac{1}{\omega - \tilde\xi_0({\bf k}) - \tilde\xi'({\bf k}) - 
\frac{|\Delta_R({\bf k})|^2}{\omega+\tilde\xi_0({\bf k})}},
\end{eqnarray}
where now
$$
\tilde\xi'({\bf k}) = -2g_t(x)t_0 + \xi'({\bf k}).
$$
Most importantly, this YRZ propagator retains a line of Luttinger zeros along the magnetic BZ
or Umklapp surface.
\end{document}